\documentclass[twocolumn,aps,floatfix,superscriptaddress]{revtex4}
\usepackage{amsmath,amssymb,graphicx}
\begin{document}
\title{Recurrence Statistics of Great Earthquakes}
\author{E.~Ben-Naim}
\affiliation{Theoretical Division and Center for Nonlinear
Studies, Los Alamos National Laboratory, Los Alamos, New Mexico
87545}
\author{E.~G.~Daub}
\affiliation{Theoretical Division and Center for Nonlinear
Studies, Los Alamos National Laboratory, Los Alamos, New Mexico
87545}
\affiliation{Earth and Environmental Sciences Division, 
Los Alamos National Laboratory, Los Alamos, New Mexico 87544}
\affiliation{Institut des Sciences de la Terre, Universit\'e Joseph Fourier, BP 53, 
Grenoble, France 38041}
\author{P.~A.~Johnson}
\affiliation{Earth and Environmental Sciences Division, 
Los Alamos National Laboratory, Los Alamos, New Mexico 87544}
\begin{abstract}
We investigate the sequence of great earthquakes over the past
century.  To examine whether the earthquake record includes temporal
clustering, we identify aftershocks and remove those from the
record. We focus on the recurrence time, defined as the time between
two consecutive earthquakes.  We study the variance in the recurrence
time and the maximal recurrence time. Using these quantities, we
compare the earthquake record with sequences of random events,
generated by numerical simulations, while systematically varying the
minimal earthquake magnitude $M_{\rm min}$.  Our analysis shows that
the earthquake record is consistent with a random process for
magnitude thresholds \hbox{$7.0\leq M_{\rm min}\leq 8.3$}, where the
number of events is larger.  Interestingly, the earthquake record
deviates from a random process at magnitude threshold \hbox{$8.4\leq
M_{\rm min}\leq 8.5$}, where the number of events is smaller; however,
this deviation is not strong enough to conclude that great earthquakes
are clustered. Overall, the findings are robust both qualitatively and
quantitatively as statistics of extreme values and moment analysis
yield remarkably similar results.
\end{abstract}
\maketitle

\section{Introduction}

Remote triggering of large earthquakes, where one large earthquake
causes another large earthquake at a global distance comparable to the
size of the earth, is the subject of ongoing debate in geophysics.  It
is well known that earthquakes do cause aftershocks on local scales,
at distances comparable to the size of the fault.  In the last 20
years, it has been shown that seismic waves can dynamically trigger
earthquakes at large distances \citep{hill,gomberg,freed}, and more
recently, that a large earthquake can trigger other large earthquakes
at global distances \citep{pollitz}.  However, other recent studies
suggest that dynamic triggering of large earthquakes is not widespread
\citep{parsons11,elst}. Thus, dynamic triggering of large events at
global distances remains an open question, one with potentially
significant implications for hazard analysis and earthquake physics.

\begin{figure}[t]
\vspace{.1in}
\includegraphics[width=0.42\textwidth]{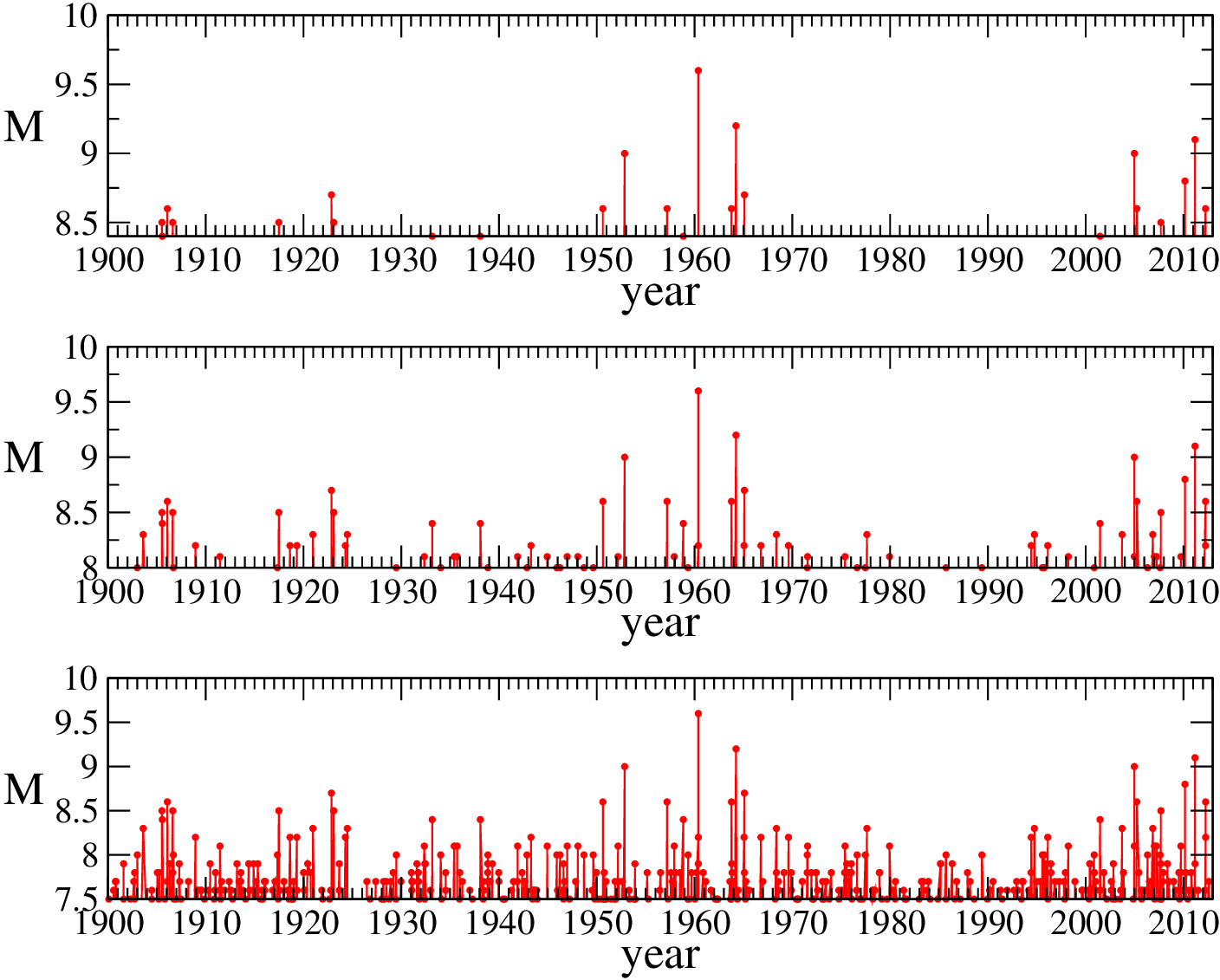}
\caption{The sequence of large earthquakes during the years
  $1900-2012$. Three thresholds are used: $M_{\rm min}=7.5$ (bottom),
  $M_{\rm min}=8.0$ (middle), and $M_{\rm min}=8.5$ (top).}
\label{fig-sequence}
\end{figure}

Remote triggering necessarily implies that large earthquakes are
correlated in time, that is, earthquakes are not equivalent to a
random process. The increase in earthquake activity over the past
decade including three of the six largest events on record over the
past century \citep{brodsky,ammon} raises the question whether great
earthquake are clustered (Fig.~\ref{fig-sequence}).

Recent studies have utilized a variety of statistical methods to
examine whether the sequence of large earthquakes is consistent with a
random process.  The approaches used to analyze the earthquake record
include for example, statistics of the number of events in a fixed
time interval, and statistics of the time between events. However, the
small number of powerful events constitutes a serious challenge for
such investigations \citep{kerr,deoliviera}. To date, some studies
reported deviations from random event statistics
\citep{buffe05,buffe11}, while several others report that the
earthquake record is consistent with random statistics
\citep{michael,shearer,parsons12,daub}.

In this study, we focus on the recurrence time between successive
earthquake events, a quantity that allows us to probe the most
powerful events on record.  Our statistical analysis quantifies
typical properties as well as extremal properties of the recurrence
time.  Using numerical simulations, we generate a large number of
random sequences, thereby allowing probabilistic comparison between
the earthquake record and a random process.

\section{Earthquake record}

We analyze the earthquake event times in the USGS PAGER (Prompt
Assessment of Global Earthquake Response) catalog \citep{allen},
supplemented by the Global CMT (Centroid Moment Tensor) catalog
\citep{ekstrom}. These two catalogs comprise a global record from 1900
through December 31, 2012, containing 1770 events with magnitude $M\geq
7$ (Table I).

\begin{table}[t]
\centering
\begin{tabular}{c r r}
\hline
$M_{\rm min}$ & All events & Aftershocks removed\\
\hline 
7.0 & 1770 & 1255 \\
7.5 &  447  & 371  \\
8.0 & 84  & 74   \\
8.5 &  19  & 17 \\
9.0 &   5  & 5  \\
9.5 &   1  & 1  \\
\hline
\end{tabular}
\caption{The number of large events on record during the years
  $1900-2012$. Listed are the total number of events (with and without
  aftershocks) versus the minimum magnitude $M_{\rm min}$.}  
\end{table}

The catalog contains aftershocks, which must be identified to address
whether earthquake occurrence is random over global distances. Removal
of aftershocks is not a trivial procedure, as it requires assumptions
that cannot be tested due to limited data \citep{marsan}.  We identify
aftershocks using a window method \citep{gardner}: any event close
enough to another larger event in both space and time is considered an
aftershock, and is removed from the catalog. We examine a variety of
choices for the distance and time windows and verify that our
conclusions are robust with respect to the aftershock removal
procedure.  In the following, we use the time window in the original
Gardner and Knopoff study, and our choice for the distance window is a
purposely conservative estimate of the rupture length for a given
magnitude (i.e. overestimated spatial extent of aftershocks), based on
an empirical law \citep{wells}.  We note that our analysis classifies
two of the $M=8.5$ events as aftershocks: the $M = 8.6$ 2005 Nias
earthquake, and the $M = 8.5$ 2007 Sumatra earthquake, both
aftershocks of the 2004 $M = 9.0$ Sumatra-Andaman earthquake.  Without
aftershocks, the catalog contains 1255 events (Table 1). For
completeness, we include in our investigation both the raw earthquake
catalog as well as the catalog with aftershocks removed.

\section{Recurrence Statistics}

The basic quantity in our analysis is the recurrence time, defined as
the time between two successive events. Recurrence times are commonly
used to characterize seismic activity.  For a random process, where
events occur at a constant rate and there are no correlations between
different events, the cumulative distribution $P(t)$ of recurrence
intervals that are larger than $t$ is purely exponential,
\begin{equation}
\label{exp}
P(t)=\exp\left(-t/\langle t\rangle\right).
\end{equation}
Here, $\langle t\rangle$ is the average recurrence time.  

\begin{figure}[t]
\includegraphics[width=0.475\textwidth]{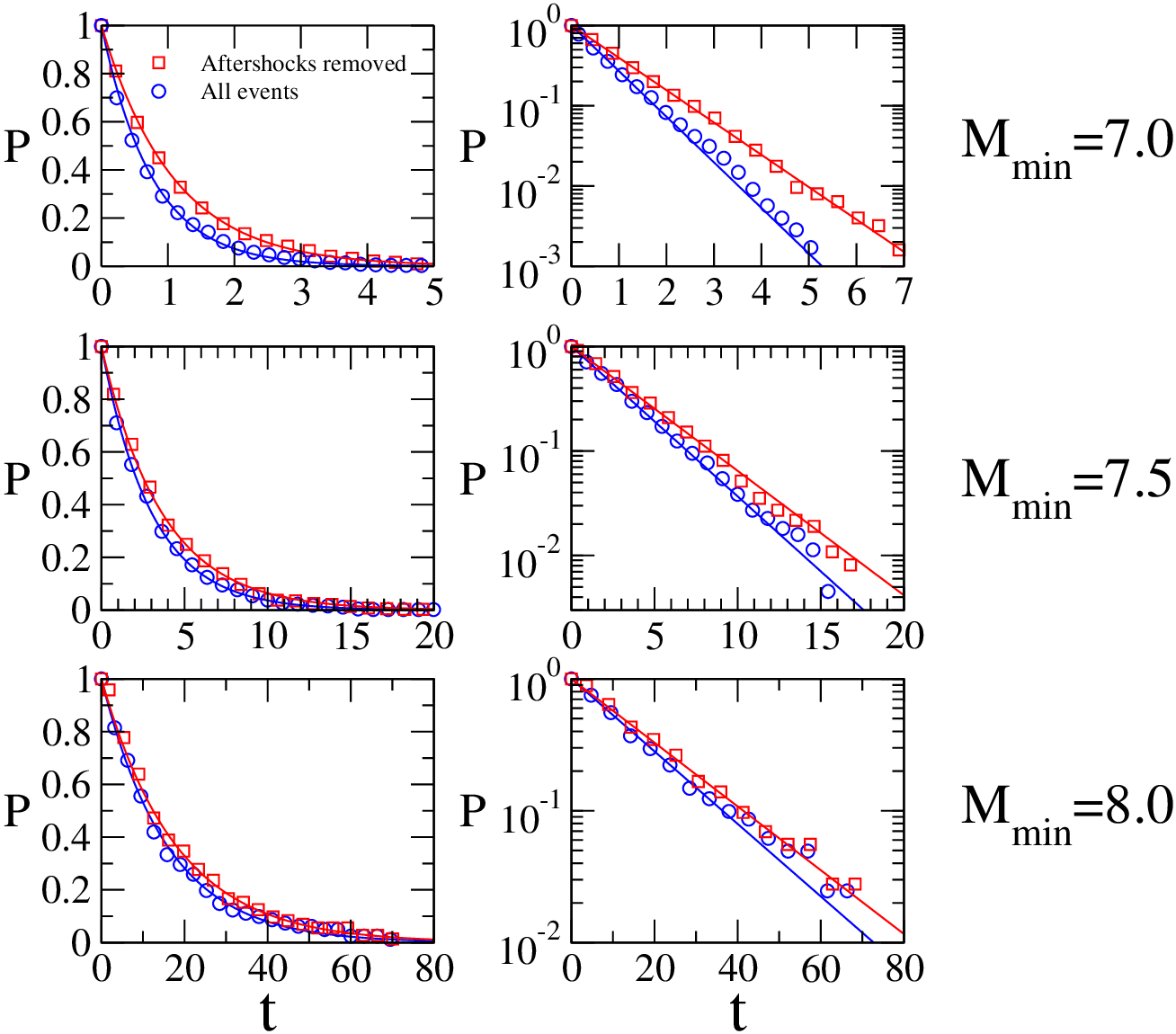}
\caption{The cumulative distribution of recurrence time versus time
  (in months). Data for all events is shown in circles and data for
  the earthquake catalog without aftershocks is shown in squares. The
  line displays an exponential distribution, where the average
  recurrence time matches the empirical value.}
\label{fig-pt}
\end{figure}

For reference, the distribution \eqref{exp} with the average $\langle
t\rangle$ calculated from the empirical data is also shown in figure
\ref{fig-pt}. The empirical distribution is in good agreement with the
exponential distribution for magnitude thresholds $M_{\rm min}=7.0$,
$M_{\rm min}=7.5$, and $M_{\rm min}=8.0$. In two cases ($M_{\rm
min}=7.0$ and $M_{\rm min}=8.0$) the tail of the distribution becomes
closer to an exponential once aftershocks are removed. This
preliminary analysis indicates that the sequence of large earthquake
events does not show significant deviations from random event
statistics.

As the magnitude threshold increases, the number of events becomes
smaller and the distribution of recurrence times can be probed only
over a smaller range.  Consequently, the tail of the distribution,
which quantifies the likelihood of large gaps between events, becomes
difficult to measure.  To address this issue and to systematically
probe high magnitudes, we analyze a standard measure for
fluctuations, the normalized variance
\begin{equation}
\label{var}
V=\frac{\langle t^2\rangle-\langle t\rangle^2 }{\langle t\rangle^2}.
\end{equation}
Here the bracket denotes an average over all recurrence intervals in
the sequence. The variance involves the lowest (nontrivial) integer
moment of the distribution, yet, as discussed below, we also analyze
a range of other moments.

We use numerical simulations to characterize how the normalized
variance behaves for a random process. We generate a very large number
($10^8$) of random sequences where the recurrence times are identical
and independently distributed variables, drawn from the exponential
distribution \eqref{exp}.  The number of events $N$ and the average
recurrence time $\langle t\rangle$ are set by the earthquake record,
for each magnitude threshold $M_{\rm min}$. By simulating the precise
number of events $N$ on record, our analysis properly quantifies the
large fluctuations that are expected when the number of events is
small.

\begin{figure}[t]
\vspace{.1in}
\includegraphics[width=0.425\textwidth]{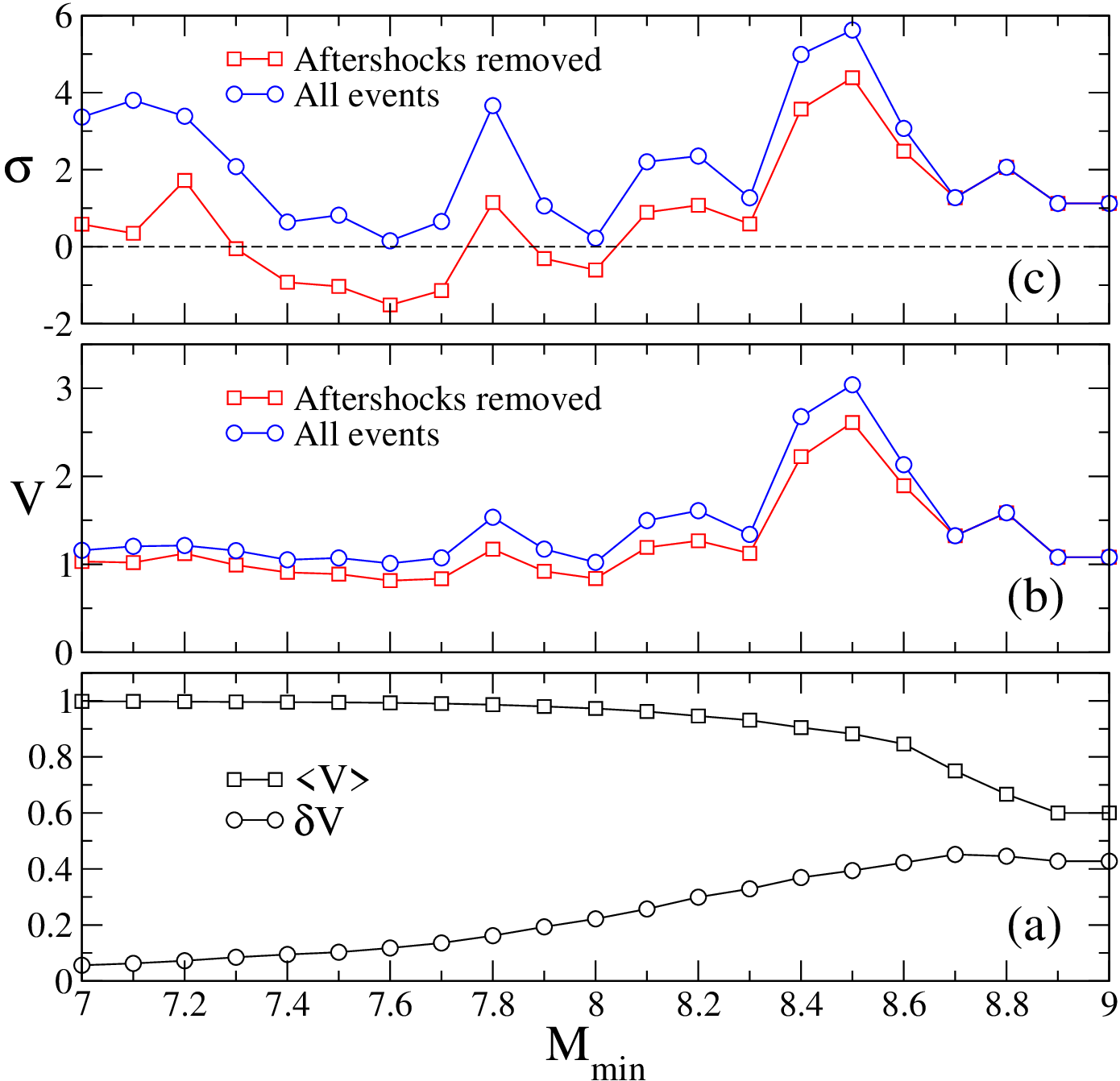}
\caption{(a) The average variance $\langle V\rangle$ and the standard
  deviation of the variance $\delta V$ as a function of magnitude
  threshold $M_{\rm min}$. These quantities correspond to a random
  sequence with a number of events that matches that of the earthquake
  record (aftershocks removed).  (b) Normalized variance $V$ as a
  function of $M_{\rm min}$. Shown are the behaviors with and without
  aftershocks. (c) The number of standard deviations away from the
  mean $\sigma$ defined in Eq.~\eqref{sigma}, versus $M_{\rm min}$.}
\label{fig-var}
\end{figure}

We measure the average variance, $\langle V\rangle$, and the standard
deviation in the variance, $\delta V$, defined by $(\delta
V)^2=\langle V^2\rangle-\langle V\rangle^2$ (here, the bracket denotes
an average over all random sequences). As shown in figure
\ref{fig-var}a, the normalized variance is close to unity when the
number of events is large, but when the number of events is small (at
large magnitudes), the expected variance decreases, and the standard
deviation becomes comparable to the mean. We also confirm that as
expected, $\delta V\sim N^{-1/2}$ for large $N$.

The normalized variance defined in Eq.~\eqref{var} is shown as a
function of the threshold magnitude in Fig.~\ref{fig-var}b.  Using the
average $\langle V\rangle$ and the standard deviation $\delta V$
obtained from simulated sequences, we also calculate $\sigma$ the
number of standard deviations away from the mean
(Fig.~\ref{fig-var}c),
\begin{equation}
\label{sigma}
\sigma=\frac{V-\langle V\rangle}{\delta V}.  
\end{equation} 
For most magnitude thresholds, even without removing aftershocks, the
quantity $\sigma$ is not large, evidence that the earthquake sequence
is consistent with a random process.  There are however three
significant peaks that indicate potential deviations from random event
statistics.  First, at the magnitude thresholds \hbox{$7.0\leq M_{\rm
min}\leq 7.2$}, the raw earthquake catalog deviates from a random
process, but once aftershocks are removed, these deviations are
largely eliminated.  Second, there is a peak at $M_{\rm min}=7.8$, but
again, this peak is eliminated once aftershocks are removed. Third,
the most pronounced peak occurs when $8.4\leq M_{\rm min}\leq 8.5$. In
this case, however, removing aftershocks diminishes the magnitude of
the peak only slightly (for such powerful earthquakes, aftershocks are
of course rare, see Table 1).

\begin{figure}[t]
\vspace{.1in}
\includegraphics[width=0.4\textwidth]{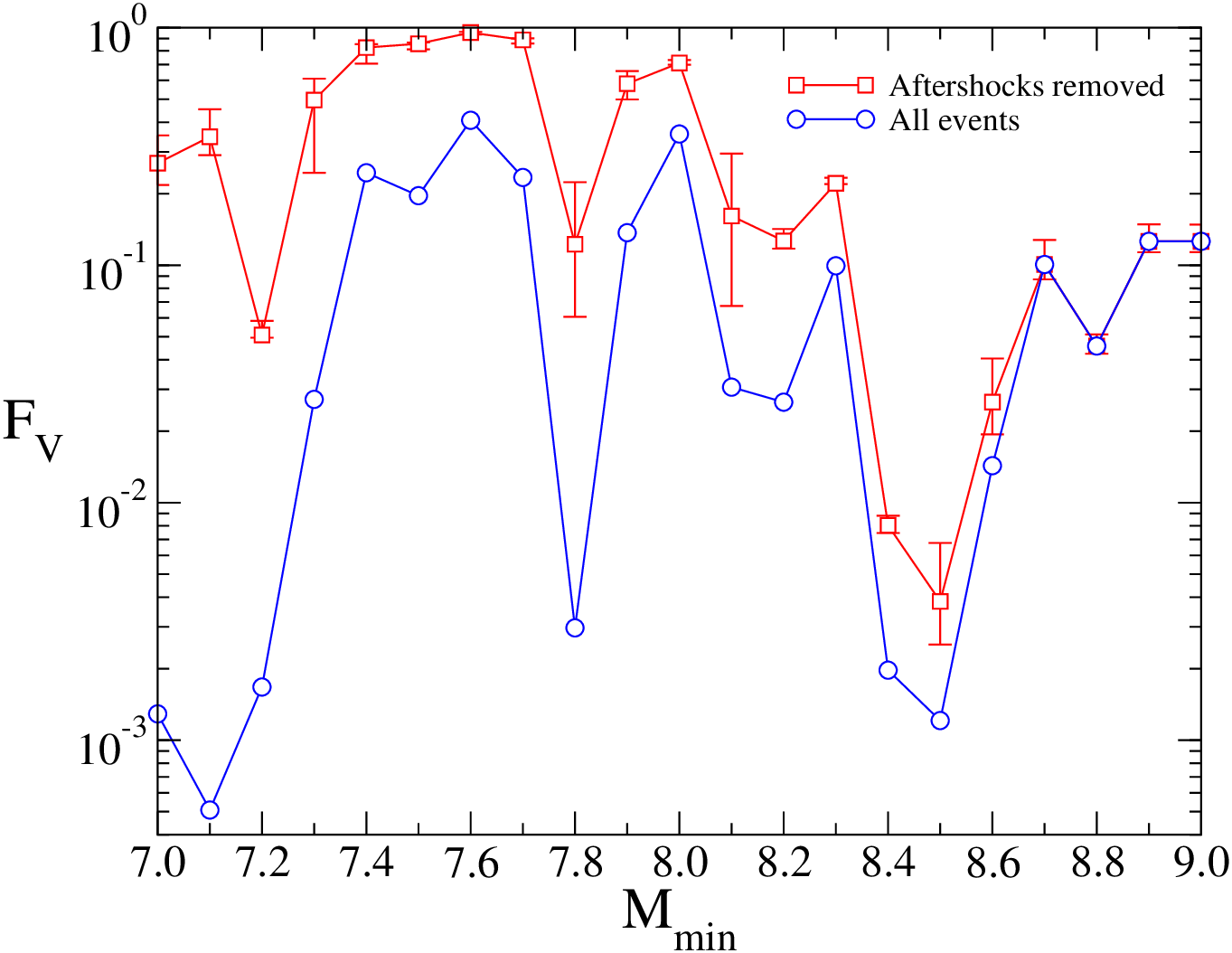}
\caption{The fraction $F_V$ of random sequences with variance
exceeding the empirical value versus magnitude threshold $M_{\rm
min}$.  Shown are results for the raw catalog (circles) and the
catalog with aftershocks removed (squares). The error bars were
produced using the moment analysis described in the main text.}
\label{fig-fv}
\end{figure}

To quantify the significance of the peaks in the quantity $\sigma$, we
use probabilistic analysis. Such analysis requires numerical
simulations because the distribution of the variance depends strongly
on the number of events: this distribution approaches a normal
distribution as the number of events becomes very large, but it is
much broader when the number of events is small. Specifically, we
measure the fraction $F_V$ of simulated random sequences where the
normalized variance exceeds the empirical value $V$. Figure
\ref{fig-fv} shows the fraction $F_V$ as a function of magnitude
threshold $M_{\rm min}$.  For each peak in $\sigma$, there is a
corresponding dip in the fraction $F_V$. These dips are mostly
suppressed once aftershocks are removed.  Yet, the dip at the narrow
band $8.4\leq M_{\rm min}\leq 8.5$ is robust. At $M_{\rm min}=8.5$ we
find $F_V\approx 1/300$, that is, only one in about $300$ random
sequences has a variance that exceeds that of the earthquake
data. This small fraction implies that the earthquake record deviates
from a random process at this particular magnitude threshold.  As
pointed out by Shearer and Stark \cite{shearer}, because $M_{\rm
min}=8.5$ is chosen a posteriori, the measured fraction $F_V$ may
represent an underestimate. Regardless, the fraction $F_V$ is not
sufficiently small to conclude with confidence that the earthquake
record violates random statistics or equivalently, that there are
temporal correlations (or causal relationships) between large events.

As a reference, our simulations show if $3$, $6$, or $9$ additional
$M\geq 8.5$ events occur over the next decade \citep{shearer}, the
quantity $F_V$ would then drop to $1.8\times 10^{-3}$, $2.9\times
10^{-4}$, and $9\times 10^{-5}$, respectively. The change in the
quantity $F_V$ with even a few additional events illustrates the
uncertainties associated with such small catalogs.

\begin{figure}[t]
\includegraphics[width=0.45\textwidth]{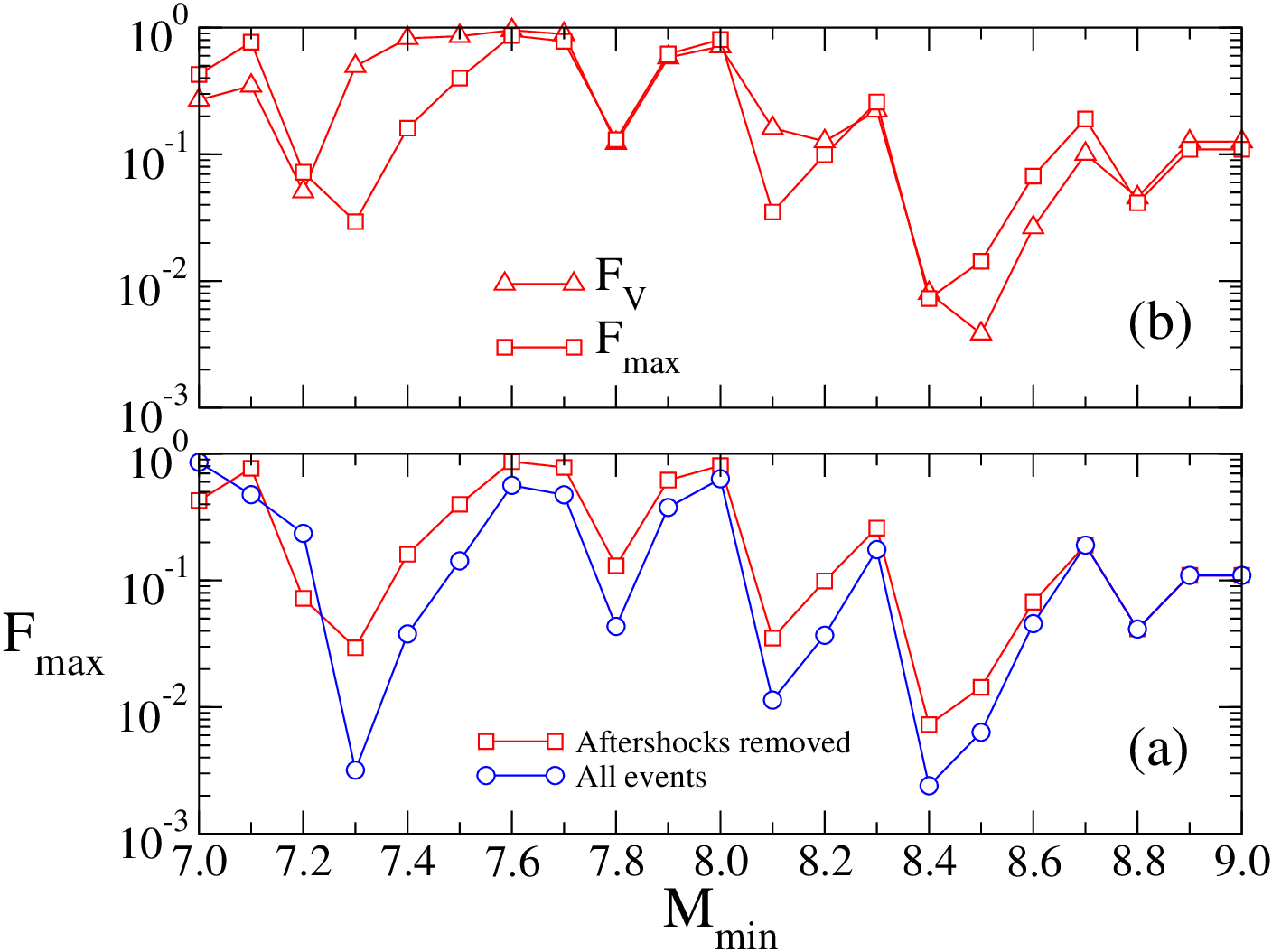}
\caption{(a) Fraction $F_{\rm max}$ of random sequences where the
maximal recurrence time exceeds the largest recurrence time on record
versus magnitude threshold $M_{\rm min}$. (b) The fraction $F_V$ (see also
figure \ref{fig-fv}) and the fraction $F_{\rm max}$ for the earthquake
catalog without aftershocks.}
\label{fig-fmax}
\end{figure}

For further insight, we examine statistical properties of the maximal
recurrence time $t_{\rm max}$, corresponding to the longest quiescent
period between consecutive earthquakes. Similar to the probabilistic
analysis above, we measure the fraction $F_{\rm max}$ of random
sequences where the maximal recurrence time exceeds $t_{\rm
max}$. When the number of events is large, this fraction is given by
the formula \hbox{$F_{\rm max}= 1-[1-\exp(-t/\langle t\rangle)]^N$}.
Fig.~\ref{fig-fmax}a shows the fraction $F_{\rm max}$ as a function of
$M_{\rm min}$. Statistics of the largest recurrence time are strongly
correlated with those of the variance: the fraction $F_{\rm max}$
mirrors the behavior of the fraction $F_V$ (Figures \ref{fig-fv} and
\ref{fig-fmax}a). Moreover, if we restrict our attention to magnitudes
$M_{\rm min}\geq 7.7$ where aftershocks are rare, the two fractions
are remarkably close to each other (figure \ref{fig-fmax}b).  Indeed,
we verify that simulated random sequences with a maximal gap that
exceeds $t_{\rm max}$ also have variance that exceeds $V$. This
extreme event analysis demonstrates that the very large $39.9$ year
gap separating two clusters of activity, one during 1950-1965 and one
during 2004-2012 is responsible for the anomalously large variability
observed at the magnitude threshold $M_{\rm min}=8.5$ (figure
\ref{fig-sequence}). 

\begin{figure}[t]
\includegraphics[width=0.45\textwidth]{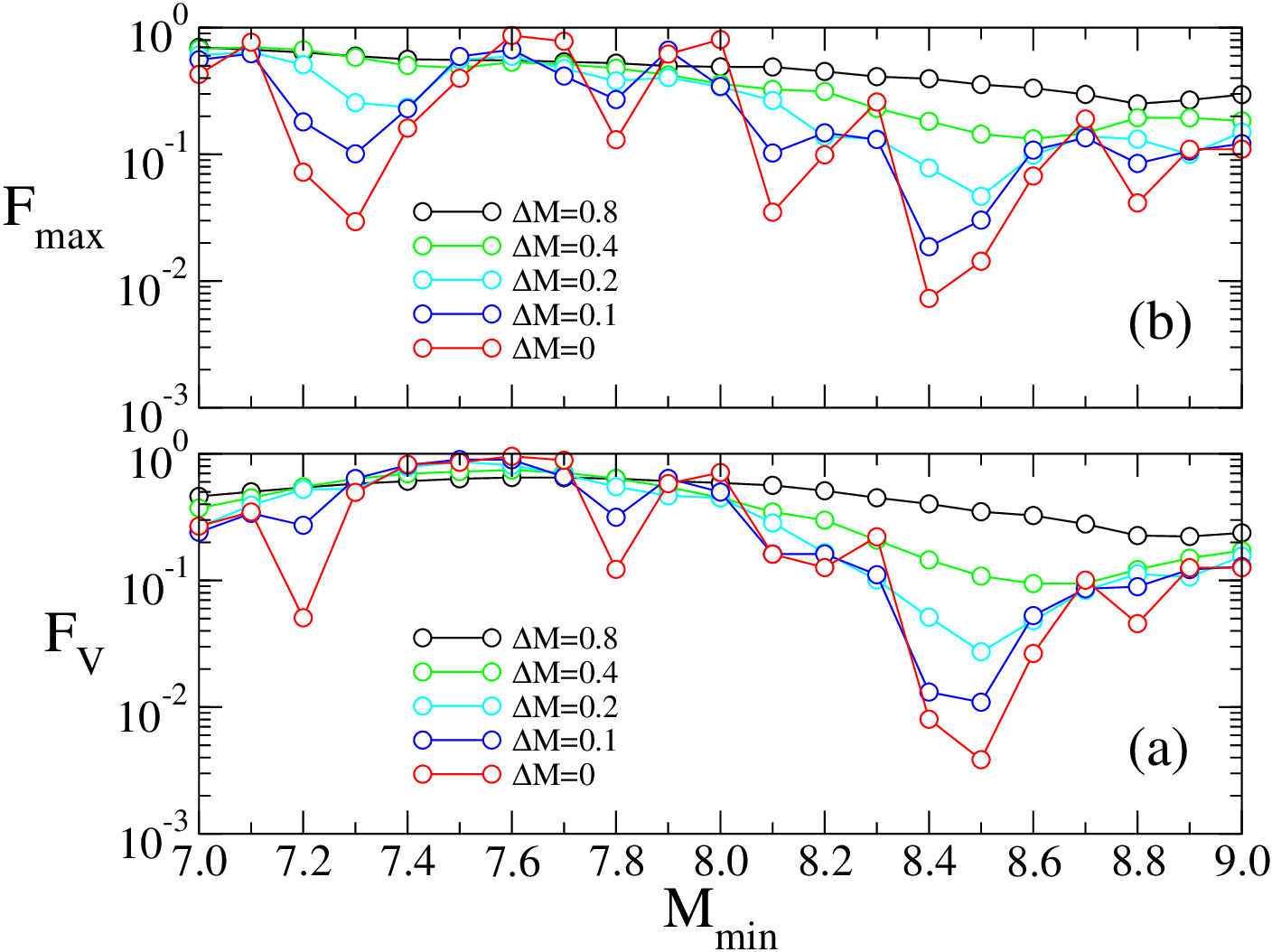}
\caption{Magnitude uncertainty analysis for the earthquake catalog
(aftershocks removed). Shown are: (a) The quantity $F_V$ as in figure
\ref{fig-fv} and (b) the quantity $F_{\rm max}$ as in figure
\ref{fig-fmax} versus $M_{\rm min}$.  The quantity $\Delta M$
quantifies the range of magnitude uncertainty, and the different
curves represent different values of $\Delta M$ in the plots.}
\label{fig-blur}
\end{figure}

Previous statistical analysis based on the number of events in a given
time interval revealed deviations from random event statistics at this
magnitude range that can be traced to magnitude uncertainties in the
earlier part of the century \citep{daub}.  To assess the effects of
uncertainties in the earthquake magnitude \citep{engdahl}, we
introduce unbiased variations in the magnitude: \hbox{$M\to M+\delta
M$} where $\delta M$ represents a potential measurement error. The
quantity $\delta M$ is drawn from a uniform distribution in the range
$[-\Delta M:\Delta M]$. We systematically increase the range $\Delta
M$ up to as high as $\Delta M=0.8$, and repeat the analysis used to
obtain figures \ref{fig-var}-\ref{fig-fmax}.  Each data point is
obtained using $10^8$ simulated catalogs: $10^4$ distinct
modifications of the original earthquakes catalog were generated, and
for each modification, $10^4$ simulated catalogs were produced.  The
fractions $F_V$ and $F_{\rm max}$ become smoother as the range $\Delta
M$ increases (Figure \ref{fig-blur}), and moreover, the dips at
$M=8.5$ are strongly suppressed. We also consider situations where the
magnitude is always underestimated or overestimated by uniformly
drawing $\delta M$ in the range $[0:\Delta M]$ or $[-\Delta
M:0]$. Biased errors lead to the same patterns shown in figure
\ref{fig-blur}.  We also verify that variations $\delta M$ drawn from
a normal distribution with standard deviation $\Delta M$ lead to
similar results.  Consistent with Ref.~\citep{parsons12}, magnitude
uncertainty analysis supports the conclusions of our statistical
analysis.

To examine whether the results are sensitive to the particular measure
of variability \eqref{var}, we repeat the analysis using the
normalized moments $M_n=\langle t^n\rangle/\langle t\rangle^n$ instead
of the variance $V=M_2-1$.  We examine a series of moments in the
range $1.25\leq n\leq 4$ and again, measure the fraction $F_M$ of
simulated catalogs where the moment $M_n$ exceeds the value measured
for the earthquake data. By varying the parameter $n$ and identifying
the maximal and minimal fractions $F_M$, we produce the error bars
shown in figure \ref{fig-fv}.  The results of this moment analysis
confirm that the dips in the quantity $F_V$ are robust. 

\section{Conclusions}

In summary, we analyzed typical and extremal properties of the time
intervals between large earthquakes. The results of our statistical
tests reconcile recent studies that address the question: ``are great
earthquakes clustered?'' Our study yields three important conclusions.

First, in the magnitude threshold range $7.0\leq M_{\rm min}\leq 8.3$
which constitutes the vast majority of great earthquakes on record,
the earthquake sequence does not exhibit significant deviations from a
random set of events.  These findings reinforce the results of several
studies \citep{michael,shearer,parsons11,daub}. At several threshold
magnitudes, the earthquake record is consistent with a random process
even if aftershocks are not removed from the catalog.

Second, the roughly twenty most powerful events on record,
corresponding to magnitude threshold \hbox{$8.4\leq M_{\rm min} \leq
8.5$}, deviate from a random sequence of events. This departure is
tied to the anomalously long gap between two clusters of events, one
in the mid-century, and one over the past decade, an observation also
noted in \citep{buffe05,buffe11,shearer}. However, this departure is
not sufficiently strong to conclude that there are temporal
correlations between great earthquakes: the likelihood that a random
sequence matches the variability in the data ($\approx 1/300$) is
equivalent to only $\approx 2.6$ standard deviations from the mean for a
normal distribution.

Third, the results are qualitatively and quantitatively robust.
Analysis of average properties and analysis of extremal properties of
the recurrence time leads not only to similar conclusions, but also,
to very similar likelihood figures that the observed sequence of
events can be explained by a random process. We have also considered
magnitude uncertainties using unbiased and biased measurement errors
in earthquake magnitude and observed that such errors systematically
suppress deviations from random event statistics.

Finally, our study uses the average recurrence time as a measure for
the overall rate of events. Uncertainties in the overall rate of
events are significant when the number of events is small, and an
important challenge for future research is to generalize the analysis
above to incorporate uncertainties in the overall rate of events.

\medskip
We thank Robert Guyer, Robert Ecke, Joan Gomberg, and Thorne Lay for
comments. We gratefully acknowledge support for this research through
DOE grant DE-AC52-06NA25396.


\begin{thebibliography}{}

\bibitem[{\textit{Hill et~al.~}(1993)}]{hill}
Hill, D. P., et al. (1993), 
Remote seismicity triggered by the M7.5 Landers, California earthquake of June 28, 1992, 
\textit{Science} \textit{260}, 1617-1623.

\bibitem[{\textit{Gomberg et~al.~}(2004)}]{gomberg}
Gomberg, J., P.~Bodin, K.~Larson, and H.~Dragert (2004), 
Earthquake nucleation by transient deformations caused by the $M=7.9$ Denali, Alaska, earthquake, 
\textit{Nature} \textit{427}, 621-624.

\bibitem[{\textit{Freed}(2005)}]{freed}
Freed, A.~M. (2005), 
Earthquake triggering by static dynamic, and postseismic stress transfer, 
\textit{Ann. Rev. Earth Plant. Sci.} {\bf 33}, 335-367.

\bibitem[{\textit{Pollitz et~al.~}(2012)}]{pollitz} 
Pollitz F.~F.~,R.~S.~Stein, V.~Sevilgen, and R.~Burgmann (2012), 
The 11 April 2012 east Indian Ocean earthquake triggered large 
aftershocks worldwide,
\textit{Nature} \textit{490}, 250.

\bibitem[{\textit{van der Elst et~al.~}(2013)}]{elst}
van der Elst, N.~J., E.~M.~Brodsky, and T.~Lay, (2013), 
Remote Triggering Not Evident Near Epicenters of Impending Great Earthquakes, 
\textit{Bull. Seismol. Soc. Am.} \textit{103}, 1522-1540.

\bibitem[{\textit{Parsons and Velasco}(2011)}]{parsons11}
Parsons, T., and A.~A.~velasco (2011), 
Absence of remotely triggered large earthquakes beyond the mainshock region, 
\textit{Nat. Geosci.} \textit{4}, 312-316.

\bibitem[{\textit{Ammon et~al.~}(2011)}]{ammon}
Ammon, C.~J., R.~C.~Aster, T.~Lay, and D.~W.~Simpson (2011), 
The Tohoku Earthquake and a 110-year Spatiotemporal Record of Global Seismic Strain Release, 
\textit{Seismol. Res. Lett.} \textit{82}, 455.


\bibitem[{\textit{Brodsky}(2009)}]{brodsky}
Brodsky, E.~E. (2009),  
The 2004-2008 Worldwide Superswarm, 
\textit{Eos. Trans. AGU, Fall Meet. Suppl.}, \textit{90}, S53B.


\bibitem[{\textit{Dimer de Oliveira}(2012)}]{deoliviera}
Dimer de Oliveira, F. (2012), 
Can we trust earthquake cluster detection tests?,
\textit{Geophys. Res. Lett.} \textit{39}, L052130. 

\bibitem[{\textit{Kerr}(2011)}]{kerr}
Kerr, R.~A. (2011), 
More Megaquakes on the Way? That Depends on Your Statistics,
\textit{Science}, \textit{332}, 411.


\bibitem[{\textit{Bufe and Perkins}(2005)}]{buffe05}
Bufe, C.~G., and D.~M.~Perkins (2005), 
Evidence for a Global Seismic-Moment Release Sequence, 
\textit{Bull. Seismol. Soc. Am.} \textit{95}, 833-843.

\bibitem[{\textit{Bufe and Perkins}(2011)}]{buffe11}
Bufe, C.~G., and D.~M.~Perkins (2011),
The 2011 Tohoku Earthquake: Resumption of Temporal Clustering of Earth's Megaquakes, 
\textit{Seismol. Res. Lett.} \textit{82}, 455. 

\bibitem[{\textit{Parsons and Geist}(2012)}]{parsons12}
Parsons, T., and E.~L. Geist (2012), 
Were Global $M\geq 8.3$ Earthquake Time Intervals Random Between 1900 and 2011?, 
\textit{Bull. Seismol. Soc. Am.} \textit{102}, 1583-1592.


\bibitem[{\textit{Daub et~al.~}(2012)}]{daub}
Daub, E.~G.~, E.~Ben-Naim, R.~A.~Guyer, and P.~A.~Johnson (2012),
Are megaquakes clustered? 
\textit{Geophys. Res. Lett.} \textit{39}, L06308.


\bibitem[{\textit{Michael}(2011)}]{michael}
Michael, A.~J. (2011), 
Random Variability Explains Apparent Global Clustering of Large Earthquakes, 
\textit{Geophys. Res. Lett.} \textit{38}, L21301.

\bibitem[{\textit{Shearer and Stark}(2012)}]{shearer}
Shearer, P.~M., and P.~B.~Stark, (2012), 
The global risk of big earthquakes has not recently increased, 
\textit{Proc. Nat. Acad. Sci.} \textit{109}(3), 717-721.

\bibitem[{\textit{Allen et~al.~}(2009)}]{allen}
Allen, T.~I., K.~Marano, P.~S.~Earle, and D.~J.~Wald (2009), 
PAGER-CAT: A composite earthquake catalogue for calibrating global fatality models, 
\textit{Seism. Res. Lett.} \textit{80}, 50-56.


\bibitem[{\textit{Ekstr\"om et~al.}(2012)}]{ekstrom}
Ekstr\"om, G., M.~Nettles, A.~M.~Dziewonski (2012), 
The global CMT project 2004-2010: Centroid-moment tensors for 13,017 earthquakes, 
\textit{Phys. Earth Planet. Int.} \textit{200-201}, 1-9.

\bibitem[{\textit{Marsan and Lenglin\'e}(2008)}]{marsan}
Marsan, D., and O.~Lenglin\'e (2008), 
Extending Earthquakes' Reach Through Cascading, 
Science 319, 1076-1079.

\bibitem[{\textit{Gardner and Knopoff}(1974)}]{gardner}
Gardner, J.~K., and L.~Knopoff, (1974), 
Is the sequence of earthquakes in Southern California, with aftershocks removed, Poissonian? 
\textit{Bull. Seismol. Soc. Am.} \text{64}, 1363-1367.

\bibitem[{\textit{Wells and Coppersmith}(1994)}]{wells}
Wells, D.~L., and K.~J.~Coppersmith (1994), 
New empirical relationships among magnitude, rupture length, rupture width, rupture area, and surface displacement, 
\textit{Bull. Seismol. Soc. Am.} \textit{84}, 1053-1069.

\bibitem[{\textit{Engdahl and Villasenor}(2002)}]{engdahl}
Engdahl, E. R., and A. Villasenor (2002), 
Global seismicity: 1900-1999, International Handbook of Earthquake and Engineering Seismology, 
Volume 81A, ISBN:0-12-440652-1, 665-690.











\end{thebibliography}
\end{document}